\documentclass[aps,pra,reprint,superscriptaddress]{revtex4-1}
\usepackage[dvips]{graphicx}
\usepackage{tikz}

\newcommand{\g}{g}

\usepackage[unicode=true,pdfusetitle,
 bookmarks=false,
 breaklinks=true,pdfborder={0 0 0},backref=false,colorlinks=true]
 {hyperref}
\hypersetup{
 linkcolor=blue, urlcolor=blue, citecolor=blue}

\begin{document}

\title{Floquet-Band Engineering of Shaken Bosonic Condensates}

\author{Brandon M. Anderson}
\affiliation{James Franck Institute, University of Chicago, Chicago, Illinois 60637, USA}
\author{Logan W. Clark}
\affiliation{James Franck Institute, University of Chicago, Chicago, Illinois 60637, USA}
\author{Jennifer Crawford}
\affiliation{James Franck Institute, University of Chicago, Chicago, Illinois 60637, USA}
\author{Andreas Glatz}
\affiliation{Department of Physics, Northern Illinois University, DeKalb, Illinois 60115, USA.}
\affiliation{Materials Science Division, Argonne National Laboratory, Argonne, Illinois 60439, USA.}
\author{Igor S. Aronson}
\affiliation{Materials Science Division, Argonne National Laboratory, Argonne, Illinois 60439, USA.}
\author{Peter Scherpelz}
\affiliation{Intelligence Community Postdoctoral Research Fellowship Program,
    Institute for Molecular Engineering, University of Chicago, Chicago,
    Illinois 60637, USA}
\author{Lei Feng}
\affiliation{James Franck Institute, University of Chicago, Chicago, Illinois 60637, USA}
\author{Cheng Chin}
\affiliation{James Franck Institute, University of Chicago, Chicago, Illinois 60637, USA}
\author{K. Levin}
\affiliation{James Franck Institute, University of Chicago, Chicago, Illinois 60637, USA}

\begin{abstract}
Optical control and manipulation of cold atoms
 has become an important topic in condensed matter.
Widely employed are optical lattice shaking experiments which allow the introduction
of artificial gauge fields,
the design of topological bandstructures,
and more general probing of quantum critical phenomena.
Here we develop new numerical
methods to simulate these periodically driven systems by implementing
lattice shaking directly. As a result we avoid the
usual assumptions associated with a simplified picture
based on Floquet dynamics.
A demonstrable success of our approach is that it yields
quantitative agreement with
experiment, including Kibble-Zurek scaling.
Importantly, we argue that because
their dynamics corresponds to
an effective non-linear Schr\"{o}dinger equation,
these particular superfluid studies
present a unique opportunity to
address how general Floquet band engineering is affected by interactions.
In particular, interactions cause instabilities at which the behavior of the system changes dramatically.
\end{abstract}

\maketitle

\section{Introduction}
Periodic shaking of optical lattices has become an important tool for controlling and
manipulating cold atom systems \cite{Eckardt2016}. 
Shaking experiments have elucidated
\cite{Parker2013,Clark2016,EckardtMott}
quantum critical phenomena, introduced novel
bandstructures \cite{Chapman,Dalibard}
and addressed Kibble-Zurek (KZ) scaling 
\cite{Kibble1976,Zurek1985,delCampo2014}. 
Just as for the analogous solid state systems \cite{Oka,Gedik1,Gedik2} the understanding of
these periodically driven systems is generally based on single particle dynamics within
Floquet band theory \cite{Tzoar}. This allows looking at dynamics in a stroboscopic
but effectively time independent fashion.
However, to create the novel phases suggested by
the Floquet band picture turns out to be complicated
\cite{Zenesini}.
Interaction effects as well as heating and dissipation arising from
periodic perturbations are also of concern \cite{Eckardt2016,Bilitewski2015,Choudhury2014,Lellouch2016}.
More generally, understanding the extent to which 
the physics of Floquet theory
is fully applicable is important not only to these atomic superfluids, but also
to their solid state analogues \cite{Oka,Gedik1,Gedik2}.

\begin{figure}[h]
\includegraphics[width=0.9\columnwidth]{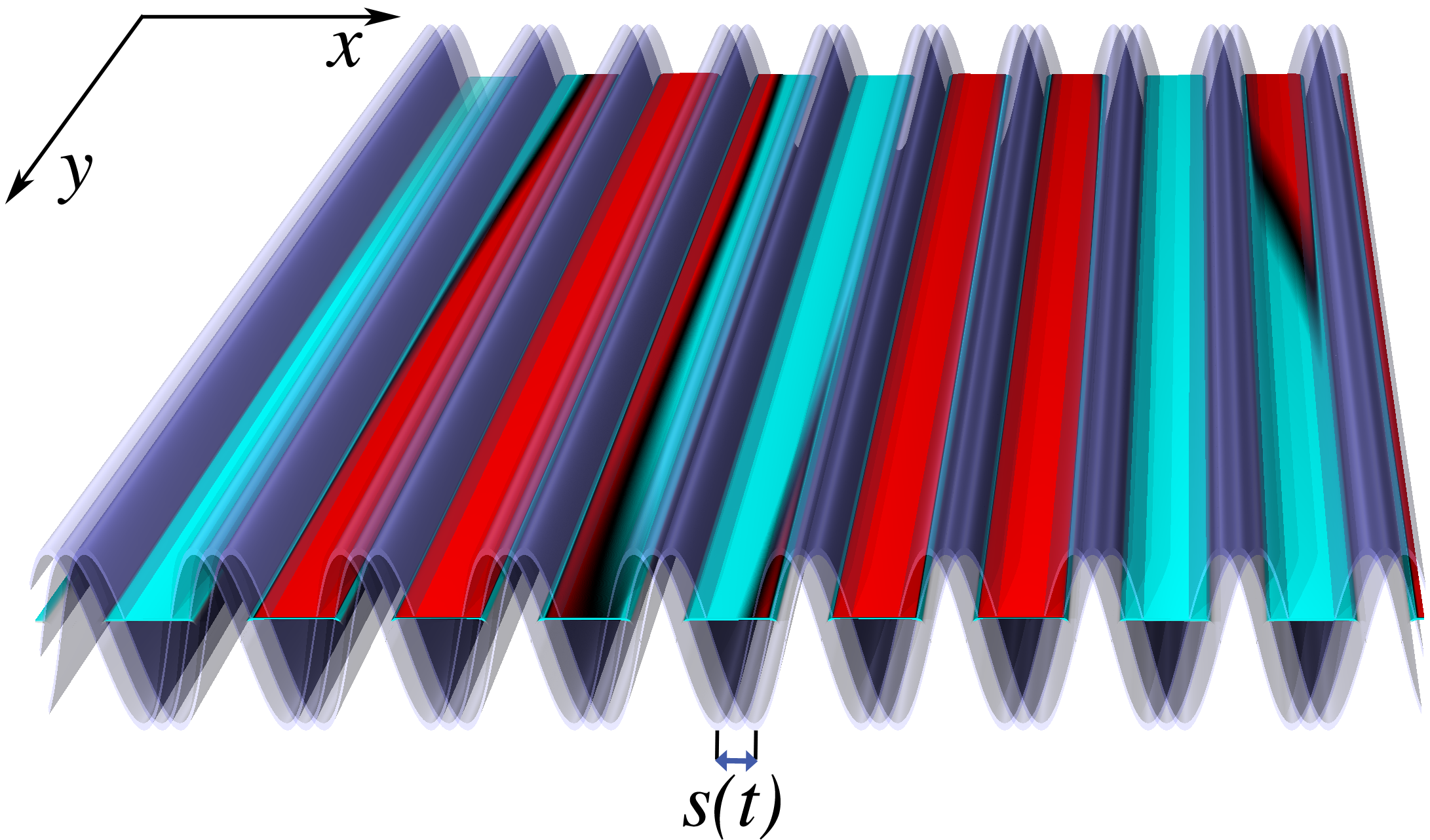}
\caption{Illustration of superfluid in shaken optical lattice as considered here. Shaking an optical lattice (purple) along the $x$-axis with a sufficient peak-to-peak amplitude $s(t)$ causes an ordinary Bose superfluid to undergo a phase transition to a finite-momentum superfluid. In this exotic phase the superfluid splits into domains with positive (red) or negative (cyan) momentum $k_x=\pm q^*$.}
\label{fig:intro}
\end{figure}

We address these issues in this paper using a microscopic
model to directly simulate unidirectional
shaking of Bose condensates.
We consider a specific (``near-resonant") frequency range~\cite{Parker2013}
and vary the interaction strengths at and substantially
away from the experimentally determined
value.
For interaction strengths comparable to experiment
\cite{Clark2016}, as was observed, we find
a quantum phase transition at fixed
shaking amplitude $s$ to an unusual multi-domain superfluid state. Importantly,
we find quantitative agreement with
these shaking experiments~\cite{Clark2016} and
show that, as in experiment, the associated domain defects are well described by
KZ scaling theory \cite{Kibble1976,Zurek1985}.
As the interaction strength, $\g$ is varied
we demonstrate how these superfluid
studies may be useful to the broader physics community. Various
instabilities appear at fixed shaking amplitudes. There is,
moreover, an interesting dichotomy, to be discussed below,
between the effects of small and large $\g$; these
are most significant in the vicinity of band crossings, and associated with non-adiabaticity.

Throughout this paper we presume a continuous linear ramp of the shaking
amplitude: $s(t) = \dot{s} t$, for $t>0$ and constant $\dot{s}$. 
We consider a shaking frequency slightly higher than the bandgap associated with the two lowest Bloch bands in the (static) optical lattice.
In this particular frequency range as $s$ varies, the Floquet dispersion
$E^{\rm floq}(k)$ transitions from having a minimum
at $k=0$ to a double well structure ($k=\pm q^*$); this occurs at a particular
critical shaking amplitude $s_c$.
This suggests that with finite interaction
strength and when $s(t)$ reaches $s_c$, a shaken Bose condensate 
will undergo a second order quantum transition to a phase
containing a mixture 
of $q^*$ and $-q^*$ domains. 

\begin{figure*}
\includegraphics[width=\textwidth]{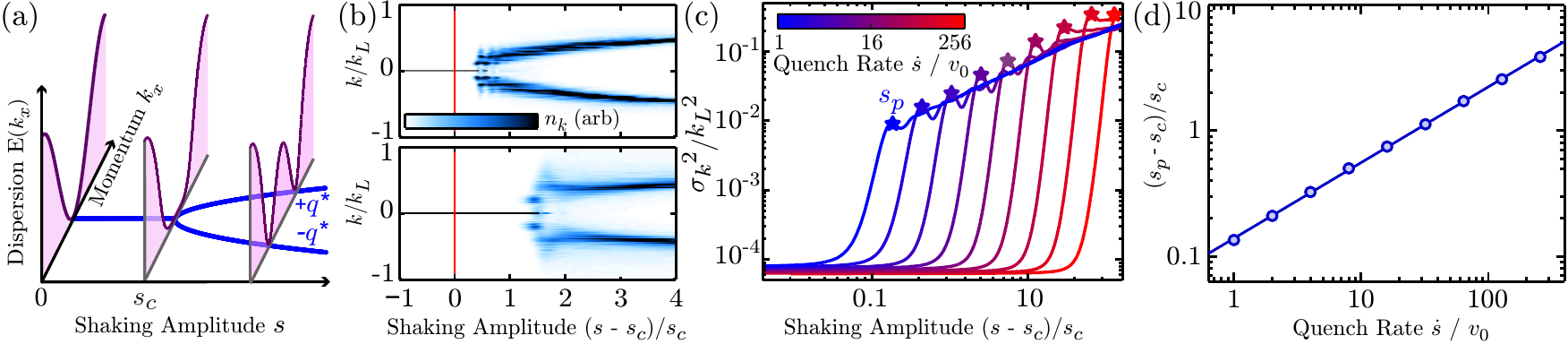}
\caption{
Dynamics of the phase transition in a shaken optical lattice.
(a) Variation in the Floquet dispersion with shaking amplitude $s$. At low shaking amplitudes the dispersion has a single minimum; at shaking amplitudes $s>s_c$ exceeding the critical shaking amplitude $s_c$, two minima appear at $\pm q^*$.
(b) Bifurcation of condensate momentum-space density, $n_k(s)$ for quench rates $\dot{s}=v_3$ (top) and $v_6$ (bottom) where $v_n=v_02^{n}$ (see Methods). Once the system gets sufficiently beyond the critical point $s_c$, the superfluid forms local domains with finite-momentum, which appears in momentum-space as a splitting of the original peak at $k=0$ into two peaks around $k=\pm q^*$.
(c) The second moment of the momentum space dispersion for a range of ramp rates $\dot{s} = v_n$, with $n=0\ldots 8$. This characterizes the amount of bifurcation, with an expected value $(q^*)^2 \propto s-s_c$ after saturation. The curves fall on this line after first overshooting to local maxima (stars), denoted by the shaking amplitude at the peak $s_p$.
(d) The shaking amplitude at the peak (circles) scales as a power-law with the quench rate $s_p\propto {\dot{s}}^{a}$ as expected from KZ theory. A fit to this equation (solid curve) yields $a \approx 0.6$.
\label{fig:k-space}
}
\end{figure*}

Figure 1 represents a schematic plot of the experimental
configuration for unidirectional lattice shaking. 
along with a sketch showing the associated $q^*$ and $-q^*$ domains 
represented by the two distinct colors. The presence of these domains was confirmed
\cite{Parker2013,Clark2016} through observation of the local current,
and an associated bifurcation in the condensate momentum-space density. 
This bifurcation, in turn, reflects the evolution to a double well structure,
as illustrated in Fig.~\ref{fig:k-space}(a).

Of central importance are dynamics as the superfluid transitions across $s_c$. 
Indeed, KZ theory~\cite{Kibble1976,Zurek1985} provides predictions for how
the ramp velocity $\dot{s}$ correlates with
the formation of topological defects, which are here associated
with domain walls. As was observed
~\cite{Clark2016}, the density of such defects increases with increasing $\dot{s}$ and 
related correlation functions were shown to be quantitatively consistent
with the scaling predictions of KZ theory.

There is extensive theoretical literature 
~\cite{Eckardt2016,Polkovnikov,ShakingTheory1,ShakingTheory2}
on the effects of shaking of optical
lattices and of KZ scaling
involving Bose Einstein condensates
\cite{GPKZ,KZdriven}. Also notable are experimental
studies which address shaking \cite{Gemelke2005,Lignier2007} and quantum
phase transitions \cite{Chapman,deMarco,BlochPNAS}, as well
as tunneling suppression and related localization \cite{dynamiclocalization}.
A numerical treatment of lattice shaking of bosonic condensates,
has, to our knowledge, not been done previously. Rather shaking is conventionally studied
\cite{Eckardt2016} by directly applying Floquet theory.
Our microscopic studies of lattice shaking go beyond Floquet
dynamics, shedding light on the engineering
of dynamic quantum states.
That our system provides good access to and agreement with the KZ
predictions derives in part from the general stability of the
domain defects. This is a consequence of the intrinsic weak
dissipation in Bose superfluids, as distinguished from their
Fermi counterparts \cite{Hannes}.
Additionally,
the robustness of the KZ scenario in the present system is likely due to 
the very strong anisotropy associated with this one dimensional
shaking of a two dimensional system.
It has been shown \cite{Aranson3,Aranson4}
that the coarsening time for 1D defects is extremely
slow, exponentially increasing with the system size.

\begin{figure*}
\includegraphics[width=\textwidth]{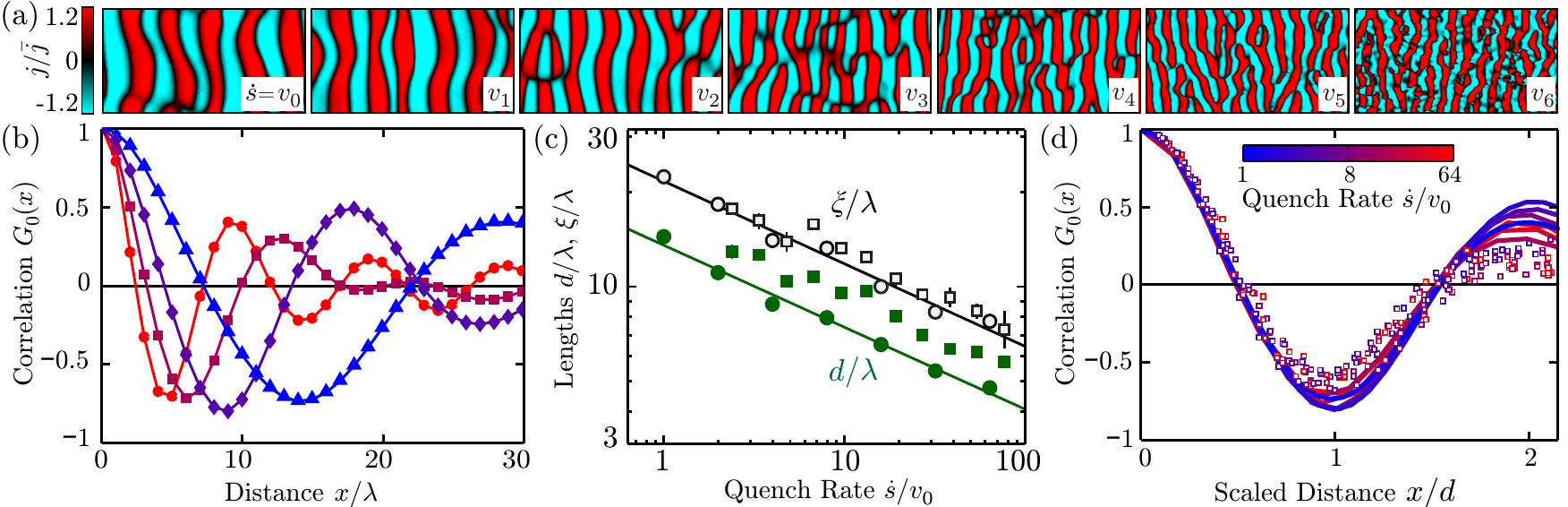}
\caption{
Comparison of spatial structure between theory and experiment.
(a) Domain structure via a snapshot of the local current density at $s=s_p$ for a range of quench rates as labelled.
(b) Plots of correlation functions $G_0(x_m)$ before rescaling the distance.
(c) Scaling of the typical domain size $d$ (solid circles) and the correlation length $\xi$ (open circles) at the shaking amplitude $s_p$ in the numerical simulations. Both length scales follow a power law scaling over a range of ramp rates as $d,\xi\propto{\dot{s}}^{-b}$ with $b \sim 0.26$. Experimental results for $d$ (solid squares) and $\xi$ (open squares) are shown for comparison \cite{Clark2016}.
(d) In scaled spatial coordinates $x/d$, the correlation functions for every quench rate collapse onto a single curve (solid lines), consistent with the experimental result (squares) \cite{Clark2016}.
\label{fig:positionscaling}
}
\end{figure*}

\section{Results and Discussion}
We study a Bose condensate in a shaken optical lattice through a full microscopic simulation of the condensate wavefunction $\psi\left(\mathbf{r},t\right)$ using the Gross-Pitaevskii Equation. We directly implement a shaken lattice potential and closely match all parameters to recent experiments \cite{Clark2016} (see Methods.) 

We are interested in physics on length scales large compared to the lattice spacing $\lambda$. 
To characterize the momentum bifurcation associated with the onset of the finite-momentum condensate,
we introduce the longitudinal momentum-space condensate density $n_{k}\left(t \right) = \int dk_y / (2\pi) |\psi(\mathbf{k},t)|^2$, where $\psi(\mathbf{k},t)$ is the momentum-space wavefunction and and we denote $k_x$ by $k$. 
From here, we can define the  second moment of the momentum-space density in the first Brillouin zone, 
$\sigma_k^2 (t) = \int_{\rm 1BZ} k^2 n_k(t) / \int_{\rm 1BZ} n_k(t)$, 
where $\textrm{1BZ} = \left[ -k_L, k_L \right]$ for $k_L = \pi/\lambda$.

We also quantify the spatial characteristics of the domain structure on length scales $x \gtrsim \lambda$. We first average the condensate wavefunction over each lattice site, $x_m$ and define $\bar{\psi}(x_m,y,t) = \int_{m\lambda}^{(m+1)\lambda} \psi(x,y,t) dx$.
We then define the site-averaged current density 
$j (x_m,y,t) = \Im \left[ \bar{\psi}^*(x_m,y,t) \delta_x \bar{\psi}(x_m,y,t) \right]$;
where $\delta_x$ is a finite difference operator. 
This leads to the normalized current-current correlation function $G_0(x) = G(x)/G(0)$, where
$G\left(x_{m}\right)=\sum_n \int j\left(x_{m}+x_{n}, y\right)j\left(x_{n}, y\right)dy$ and $G(0) = \bar{j}^2$ is the mean-squared current.

\textit{Numerical results and analysis of GPE dynamics.--}
We now present the results of our numerical simulation. We begin with the momentum-space condensate density $n_k(s)$, as shown in Fig.~\ref{fig:k-space}(b), at two different ramp rates; we henceforth express time in units of shaking amplitude through $s(t) = \dot{s} t$.
For both ramp rates the condensate density is peaked at zero momentum until some time after the expected critical shaking amplitude (defined below and marked by the vertical line) is passed. Beyond this critical amplitude, the condensate bifurcates reflecting the transition in the single
particle Floquet dispersion. Our numerical calculations lead to results very similar to the observations in Refs.~\cite{Clark2016,Parker2013}.

To quantify the physics in more
detail, in Fig.~\ref{fig:k-space}(c), we present calculations of $\sigma_k^2 (s) $ for a range of ramp rates. 
All curves reach an equilibrium value of $\sigma_k^2 (s)$ which lies on a curve consistent with an expected dependence $\sigma_k^2 \propto \left(q^*\right)^2 \propto s-s_c$. Two key features appear as the ramp rate increases: (1) For
faster ramps there is a delay in the bifurcation onset, relative to the critical point. (2) Additionally, we observe an overshoot (and sometimes a subsequent undershoot) in the curves before they settle down to the equilibrium value of $\left(q^*\right)^2$. This overshoot or ``ringing'' effect is more 
apparent the faster the ramp. Similar overshoots were also observed \cite{Aranson1}
in theoretical studies of rapid thermal quenches of He$^3$.
By studying time correlation plots from one site to another 
we establish that the overshoot stems from a collective oscillation of the entire condensate (and domain walls), with adjacent domains moving out of phase.

As can be seen,
the most easily quantified feature of Fig.~\ref{fig:k-space}(c) is the amplitude, denoted $s_p$, of the first overshoot peak. We use this peak to 
characterize the temporal scaling of the bifurcation onset. 
It should be noted that in this one regard our analysis is different from experiment \cite{Clark2016}, as we (and experiments) find 
the overshoot essentially disappears for a trapped gas.
In Fig.~\ref{fig:k-space}(d), we plot the shaking amplitude associated with
the peak versus the ramp rate on a log-log plot. The clear linear dependence of this relation suggests a power law scaling, with a corresponding exponent of $a\approx 0.6$. This can be compared with the experimental exponent of $a \approx 0.5$ \cite{Clark2016}. 
In this way, as in somewhat different contexts~\cite{tdglKZ,GPKZ,KZdriven} our
numerical simulations produce a KZ scaling over a large range of ramp rates.

Establishing an effective time scale variable then enables an analysis of position space scaling.  Here we are able to quantitatively compare
with experiment, as trapping effects appear less significant.
Figure~\ref{fig:positionscaling}(a) shows sample domain configurations near a shaking amplitude $s_p$, for a range of ramp rates. A form of self-similarity in the curves is evident with the fine domain structure shown on the right (corresponding to the most rapid ramps) and the coarser domain structure on the left (slowest ramps).  This appears qualitatively consistent
with the KZ picture. 

For a more quantitative analysis we study the spatial correlation function $G_0(x_{m},s_p)$ defined above, as shown in Fig.~\ref{fig:positionscaling}(b). 
The Fourier transform of this quantity has a characteristic peak $k_{\rm peak}$ and full-width-half-maximum, $w_k$, from which we define length scales $d\equiv \pi/k_{\rm peak}$ and $\xi\equiv\pi / w_{k}$. These length scales are functions of
$\dot{s}$ and are presented in Fig.~\ref{fig:positionscaling}(c) which shows
a range of ramp rates; both scale as $d,~\xi \propto{\dot{s}}^{-b}$ with $b\sim0.26$. This yields quantitative
agreement with experiment (shown as squares)~\cite{Clark2016}.
Finally, Fig.~\ref{fig:positionscaling}(d) presents the same correlation functions in Fig.~\ref{fig:positionscaling}(b), but with the position coordinate expressed in terms of a scaled distance $d(\dot{s})$. All correlation functions lie along a universal form.
Importantly, this universal curve
appears quantitatively consistent with the experimental data (shown as squares)~\cite{Clark2016}.

\textit{Beyond Floquet dynamics.--}
When the interaction strength assumes the experimental value, $\g = g_{\rm expt}$, 
the numerical results presented above show a consistency with experiments.
However, Floquet theory is not rigorously defined for a condensate
when many body effects become significant.
Nor is it straightforward to anticipate what happens at very much smaller
$\g$. 
Clearly, interactions cannot be neglected as
they are necessary to locally equilibrate the condensate into a unique $\pm q^*$ state. Thus, understanding the interplay of interactions and single particle Floquet physics is a subtle but crucial
challenge for engineering novel physics with periodically driven superfluids.

\begin{figure}[h]
\includegraphics[width=\columnwidth]{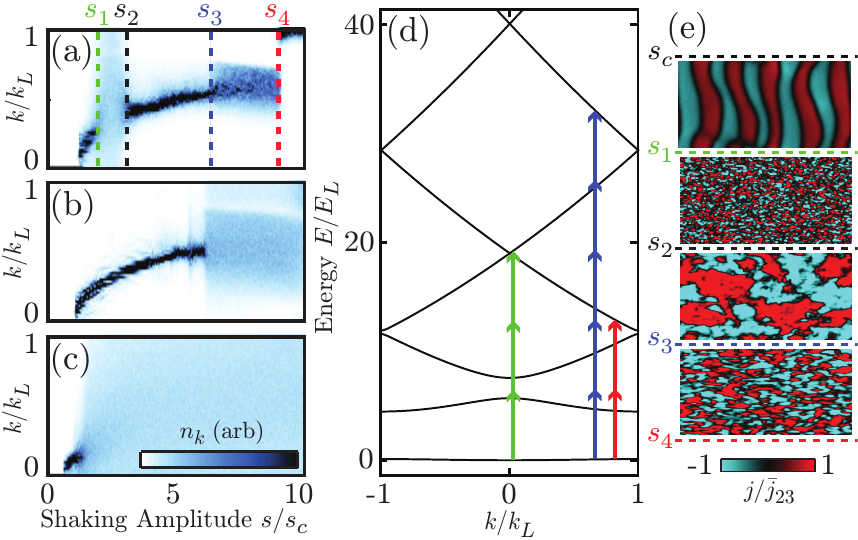}
\caption{Effect of interactions on Floquet evolution.
Panels (a), (b), and (c), show the evolution of the condensate density in momentum-space between $0<k<k_L$ for interaction strengths $\g/g_\mathrm{expt}=0.1,1,10$, respectively, for the quench rate $\dot{s} = v_{0}$. A dark, localized density indicates a region of stable domain formation; a lighter smeared density distribution indicates the loss of stable domain structure. For the weakest interactions shown in panel (a), we mark the shaking amplitudes $s_1$, $s_3$, and $s_4$ where domains are destabilized and $s_2$ where domains are stabilized.
(d)
The lowest 6 Bloch bands of the unshaken lattice system, with arrows
indicating resonant transitions which are expected to be responsible for destabilizing the domain structure. Each arrow has a length equal to the shaking frequency. 
(e) Typical current densities for a quench with $\g/g_\mathrm{expt}=0.1$ taken between the marked shaking amplitudes. Before $s_c$ and beyond $s_4$ the current density is extremely small. }
\label{fig:4}
\end{figure}

To explore this we consider a range of interaction strengths 
$\g/g_{\rm expt}$ of
$0.1$, $1$, and $10$ which vary roughly from about a factor of 10 smaller
to a factor of 10 larger than in experiment. 
These are illustrated in
Figures~\ref{fig:4}(a), \ref{fig:4}(b), and \ref{fig:4}(c) respectively,
where we present
$n_k(s)$.
For definiteness, we take 
a ramp rate $\dot{s} = v_{0}$, about a factor of two slower than explored in experiment \cite{Clark2016}.
A dark, localized density, represents a region of stable domain formation as 
reflected in the current-current correlations; a lighter smeared distribution represents the loss of localized finite-momentum domains. 
One can see that the effects of varying $\g$ are less apparent at the shortest times where domains are stable in all three cases.

Interestingly, one can identify characteristic times (or equivalently shaking
amplitudes) $s_1$ and $s_3$ which are evident not only for the weakest
interaction case of
Fig.~\ref{fig:4}(a) 
but also appear either in 
\ref{fig:4}(b) or \ref{fig:4}(c).
These features are points at which the domain structure abruptly changes; either
they disappear altogether or abruptly become incoherent.
Since one sees these transitions at characteristic amplitudes, $s_i$,
even for different interaction strengths, one might expect
that they are intrinsic to the underlying Floquet bandstructure. 
Figure \ref{fig:4}(d)  addresses this issue by presenting  
the six lowest Bloch bands of the unshaken, non-interacting system. These bands are coupled in Floquet theory when the energy difference coincides with with $n \hbar \omega$. 

We see resonant transitions from the first band are possible to the fourth, fifth, or six band (corresponding to $s_4$, $s_1$, and $s_3$ respectively). A more precise Floquet calculation predicts these transitions happen at momenta near $k \sim 0.77 k_L$, $k \sim 0.09 k_L$, and $k \sim 0.63 k_L$ respectively.
The widest crossing occurs at $s_4$, and can be seen as an abrupt transfer of the condensate to the zone edge. The crossings at $s_1$ and $s_3$ are narrower and their effects are more subtle.
For completeness in Fig.~\ref{fig:4}(e) we present the domain
configurations between each of the characteristic times $s_1$, $s_3$ and
$s_4$ that correspond to these band resonances for the
weakest interaction case. (We find that $s_2$ is not related to these
resonances, but rather appears to reflect dissipation, and is highly
sensitive to $\g$.) The figure graphically shows
that when the domains appear they are initially rather stripe like, then they
lose all spatial extent
and then begin to enlarge having shorter range
correlations in the transverse direction. Finally they disappear
as the system transitions to a single $q^*$ condensate at the zone edge.

Because the notable features in
the plots of Figs.~\ref{fig:4}(a)-(c) appear to be related to
the Floquet transitions of \ref{fig:4}(d), these can be thought
of as band crossings which represent points where adiabatic flow will break down.
We note that the effects of varying $\g$ seem to be very different in
the weak and strong interaction regimes. For the weakest interactions,
(\ref{fig:4}(a)) domains are formed for a very short time after which they are lost
at $s_1$, and then reformed at $s_2$. Importantly, a
small increase by a factor of three in $\g$ will re-instate domains in the intermediate region. 
It thus appears at weak $\g$
that one needs a small amount of inter-boson
interactions which act to assist in surpassing a level crossing barrier, where
adiabaticity breaks down.

For large $\g$ at later times,  (Fig.~\ref{fig:4}(c)),  one sees
a rather dramatic shut-down of domain stability which is
as abrupt in time as is the initial bifurcation onset for
domain formation.
It appears that the role of large and moderate $g$ is, then, to
introduce, a more turbulent behavior which makes it
difficult to surpass these band crossings or resonances of
\ref{fig:4}(d).

\section{Conclusions}
In this paper we
have focused on a very useful prototype system:  the
near-resonant unidirectional shaking
\cite{Parker2013,Clark2016}
of a bosonic condensate on an optical lattice which is treated fully
microscopically. As the shaking amplitude
is ramped up a special class of defects (domain walls) form.
Because of their one dimensionality, they are anomalously slow to heal
\cite{Aranson3,Aranson4}. This
makes them
ideal for addressing Kibble Zurek scaling which we do here
through a successful comparison of our numerical simulations and
experiment \cite{Clark2016}.

We note that this multi-domain superfluid
which results from a shaken lattice leads to a very
unusual superfluid which is associated with condensation
at finite momentum. We speculate that this may have features in common
with
a highly disordered form of the Larkin Ovchinnikov \cite{FFLO} fermionic
superfluid phase.
It appears clear that many
of its properties remain to
be characterized, and this should stimulate further lines of
research.

A highlight of this paper has been to address the combined effects of
driven periodicity and many body interactions.
Here we find that inter-Floquet-band crossings underlie much of
many body behavior, as they present themselves as
barriers
for adiabatic evolution.
Understanding these superfluid dynamics
requires the solution of
an effective non-linear dynamical Schr\"{o}dinger equation; this
thereby, provides an ideal system to
address the interplay of Floquet band engineering and interaction
effects.
Indeed, the strongest impact of our work relates to this
rather unique opportunity
which, through numerical simulation, reveals
the presence of a variety of
instabilities appearing in periodically driven systems.
Quite generally, our results should shed new light on the engineering of
dynamic quantum states.

\section{Methods}
Our Bose condensate in an optical lattice is described by the usual
mean-field Hamiltonian, with dynamics given by the
time-dependent Gross-Pitaevskii equation (GPE)
(or equivalently time dependent Landau-Ginzburg theory):
\begin{eqnarray}
i \hbar \partial_{t}\psi\left(\mathbf{r},t\right)&=&  e^{i \gamma} \left(-\hbar^2 \nabla^{2}/2m-\mu+V_{L}\left(x-\phi\left(t\right)\right) \right. \nonumber \\
& & \quad + \left. \g \left|\psi\left(\mathbf{r},t\right)\right|^{2}\right) \psi\left(\mathbf{r},t\right).
\label{eq:1a}
\end{eqnarray}
where $\psi\left(\mathbf{r},t\right)$ is the two-dimensional condensate wavefunction at time $t$ and position $\mathbf{r} = (x,y)$, $\nabla$ is the gradient operator, and $m$ is the mass of a single atom. The chemical potential $\mu$ includes zero point lattice energy $\epsilon_0$, and we rescale $\psi$ at $t=0$ so that 
$\int |\psi\left(\mathbf{r},0\right)|^4 d\mathbf{r}/ \int |\psi\left(\mathbf{r},0\right)|^2 d\mathbf{r}= 1$,
and the interaction constant $\g = \mu - \epsilon_0$ simultaneously defines an intrinsic mean-field interaction energy that can be compared to experiment~\cite{Clark2016}, where $g=g_{\rm expt} \approx 0.18 E_L$.
To focus on the central physics, we consider a homogeneous system with periodic boundary conditions. This avoids complexities with the trapping potentials in the counterpart experiments~\cite{Parker2013,Clark2016}.
Unless indicated otherwise, we find little significant changes when we include trap effects.

The one-dimensional lattice potential $V_{L}\left(x\right)=U_{0}\sin^{2}\left(k_L x\right)$ has a depth $U_{0}$, lattice constant $\lambda=\pi/k_L$ and lattice recoil energy $E_L = \hbar^2 k_L^2 / 2m$. Shaking is implemented through the time-dependent shift
$\phi\left(t\right)=s(t)/2 \sin\omega t$ of the lattice from its equilibrium value at $t=0$.
We closely match experimental parameters~\cite{Clark2016} with $U_0=8.86 E_L$, $\omega=6.04 E_L$. We chose logarithmically spaced ramp rates of the form $\dot{s} = v_0 2^n$, with $n=0 \ldots 8$ and $v_0 = 0.98 \times s_c \omega / ( 2^8 50 \pi)$; 
the range $n=1\ldots 6$ then closely matches the range of experimental ramp rates \cite{Clark2016} relevant for Figs.~\ref{fig:positionscaling}(c,d).

This GPE approach with a phenomenological dissipation parameter $\gamma > 0$
has applied with some success in a somewhat different
context~\cite{Scherpelz}.
For definiteness,
in our figures we present results for $\gamma \approx 0.01$. We also consider a small random noise introduced at each time step which is necessary to seed domain formation, and noise averaged over $50$ implementations of noise.
We have systematically checked that our general quantitative results are robust to a wide range of noise and dissipation parameters. 
When compared with alternative approaches~\cite{Blakie} which introduce
microscopic dissipation via a bath, we note that the presence of shaking will modify the dynamics of a condensate bath adding additional complexities.

Here we use a GPE solver based on a split-step
algorithm, implemented on graphic processing units.
We are able to implement noise averaged shaking simulations
for large size, $L_x = 2 L_y = 128 \lambda$, systems in a reasonable time frame. We use a grid spacing of $4096 \times 64$ points; this fully resolves each longitudinal lattice site with $32$ grid points. Our timestep, $dt = 2\pi/(100 \omega)$, resolves every shaking cycle with a hundred time slices.  
In order to capture the expected Floquet physics, observables of interest were calculated at stroboscopic time slices $\omega t=2\pi (n+1/4)$ for integer $n$.

\textit{Correlation functions and related observables.--}
KZ theory \cite{Kibble1976,Zurek1985,delCampo2014}
establishes that a general observable $f$ should be invariant in scaled space-time coordinates and therefore can be expressed in terms of a universal function $F$ as \cite{Kolodrubetz2012,Chandran2012,Francuz2016}:
\begin{equation} 
f(x,t; \dot{s}) = \dot{s}^l F(x/x_\mathrm{KZ},t/t_\mathrm{KZ})
\end{equation}

\noindent where $l$ is the scaling dimension of $f$ and the KZ length $x_\mathrm{KZ}(\dot{s})\propto\dot{s}^{-b}$ and time $t_\mathrm{KZ}\propto\dot{s}^{-a}$ exhibit a power law dependence on the quench rate $\dot{s}$.
We omit the dependence on the $y$ spatial component which was not
addressed experimentally and presumably may introduce additional
scaling exponents.
The exponents $a=z\nu/(1+z\nu)$ and $b=\nu/(1+z\nu)$ are directly related to the equilibrium spatial critical exponent $\nu$ and dynamical critical exponent $z$ which characterize the phase transition.

\textit{Acknowledgements.--}
We thank C. Parker, R. Boyack C.-T Wu for helpful conversations. This work was supported by NSF-DMR-MRSEC 1420709.
The work of I.S.A. and A.G. was supported by the US Department of Energy, Office of Science, Basic Energy Sciences, Materials Science and Engineering Division. 
L.W.C. was supported by a Grainger Graduate Fellowship.
This research was supported by an appointment (P.S.) to the Intelligence
    Community Postdoctoral Research Fellowship Program at The University of
    Chicago, administered by Oak Ridge Institute for Science and Education
    through an interagency agreement between the U.S. Department of Energy and
    the Office of the Director of National Intelligence.
    
\bibliography{Floquet2}

\end{document}